\documentclass[%
 reprint,
 amsmath,amssymb,
 aps,
pra,
]{revtex4-2}

\usepackage{graphicx}% Include figure files
\usepackage{dcolumn}% Align table columns on decimal point
\usepackage{bm}% bold math
\usepackage{xcolor}

\usepackage{ulem}

\begin{document}

\preprint{APS/123-QED}

\title{Optical neuromorphic computing based on  chaotic frequency combs in nonlinear microresonators}% Force line breaks with \\
%\thanks{\dagger footnote to the article title}%
\author{Negar Shaabani Shishavan$^{1}$} 
%\email{nshaa22@aston.ac.uk}
\author{Egor Manuylovich$^{1}$}
%\email{e.manuylovich@aston.ac.uk}
\author{Morteza Kamalian-Kopae$^2$}
\author{Auro M. Perego$^{1,}$} 
\email{a.perego1@aston.ac.uk}
\affiliation{$^1$
 Aston Institute of Photonic Technologies, Aston University, Birmingham, B4 7ET, United Kingdom}
 \affiliation{$^2$ Microsoft Azure Fibre, Romsey Quadrangle-Abbey Park, SO51 9DL, Ramsey, UK}
%\affiliation{%


%\affiliation{
%\affiliation{
 %Third institution, the second for Charlie Author
%\affiliation{%


\date{\today}% It is always \today, today,

\begin{abstract}
%Reservoir computing is one of the most promising neuromorphic computational architectures due to its potential implementation using different physical hardware and its agile training procedure. 
In this work we present a novel implementation of delay line free reservoir computing based on state-of-the-art photonic technologies, which exploits chaotic optical frequency comb formation in optical microresonator as the nonlinear reservoir. Our solution leverages the high resonator $Q$-factor both for memory and for enhancing high dimensional nonlinear mapping of input symbols.
We numerically demonstrate the accurate prediction of about one thousand symbols in chaotic time series without the need of dedicated optimisation for specific tasks. Our results will enable design of optical neuromorphic computing architectures combining on-chip integrability, low footprint, high speed and low power consumption.
%\begin{description}
%\end{description}
\end{abstract}

\maketitle

%\tableofcontents

%\section{\label{sec:level1}Introduction}
%Neuromorphic computing systems,  are computational paradigms designed to mimic human brain's functionalities, offering a promising alternative to traditional Von Neumann architectures on which exisiting computers are based. 
%While Von Neumann computational systems face physical constraints that limit computational growth, neuromorphic systems leverage brain-inspired designs where memory and processing units are not physically separated and are promising solutions for overcoming limitations of traditional dictated by Moore's law increasing computational speed with low power consumption\cite{yadav2023neuromorphic}.
%Among the diverse range of hardware used for neuromorphic computing, photonic based platforms have been attracting a substantial interest due to several potential benefits including low latency, high bandwidth, fast reconfigurability, and low power consumption \cite{ReviewOC,ShastriReview}.

Optical reservoir computing (RC) is a computational approach which leverages potential benefits of photonic platforms, such as low latency, high bandwidth, fast reconfigurability, and low power consumption \cite{ReviewOC,ShastriReview}, to perform computation in a neuromorphic fashion going beyond the traditional Von Neumann paradigm. RC is a neuromorphic computing approach within the recurrent neural networks (RNNs) family \cite{tanaka2019recent}. Its operational principle relies on a mapping of the input signal into a higher dimensional space exploiting the nonlinear dynamics of a reservoir. By training only the output layer, without changing the inner layers parameters, RC significantly reduces training time compared to standard RNN, making it an efficient and scalable solution for various computational tasks, including time-series prediction, pattern recognition, and nonlinear signal processing.

%RC computing has been successfully demonstrated so far in a diverse range of physical systems \cite{tanaka2019recent} including spintronics, fluids, electronic circuits, mechanical systems, and optical devices \cite{van2017advances, wu2022optical}.

%Optical computing, especially using photonic integrated circuits (PICs), addresses the growing demand for computational power with high bandwidth and low power consumption\cite{hochberg2013silicon, dabos2022neuromorphic}. Integrating photonic components enhances energy efficiency and processing speed, meeting the need for sustainable computing \cite{kachris2012survey}. Photonic neuromorphic information processing demonstrates robustness against fabrication tolerances and suitability for hardware implementations within RC, highlighting its potential in advancing neuromorphic computing technologies \cite{Lugnan2020}.
Photonics solutions for RC based on wave dynamics have been proposed \cite{Marcucci,Sunada,Linear} and have been implemented in several platforms including optical fibers \cite{Vinckier:15,Supercontinuum}, optical parametric oscillators \cite{OPO,OPO2}, saturable absorbers \cite{Dejonckheere:14}, media exhibiting quadratic nonlinearity \cite{Wright}, disordered scattering media \cite{Scattering} semiconductor optical amplifiers \cite{Duport:12,Vandoorne:08}, semiconductor laser arrays \cite{Brunner:15}, phase-change materials \cite{PCM},
and optical cavities and resonators \cite{Vinckier:15,
Donati:22, borghi2021reservoir,staffoli2023nonlinear,bazzanella2022microring,phang2020chaotic,GironCastro:24}. Among photonic hardware for optical RC, microresonators (MR) are promising due to compact dimensions, possible on-chip integrability, CMOS compatibility, and low power consumption through nonlinear response achievable with input power in the mW order, enabled by high-quality ($Q$) factor platforms.
In particular, so far, both theoretical and experimental investigation of RC in semiconductor MR have been performed leveraging two photon absorption and free carrier dispersion as a nonlinear tools for achieving classification and prediction tasks \cite{borghi2021reservoir,staffoli2023nonlinear, phang2020chaotic, bazzanella2022microring,Donati:22,GironCastro:24,castro2023multi}.
While several approaches to optical RC use delay lines -- for instance in MR or in other settings as well -- to enable system memory and symbol interaction \cite{Larger,borghi2021reservoir,Donati:22,castro2023multi}, in order to reduce system complexity and footprint, a particularly appealing emerging trend is the delay line free reservoir computer\cite{borghi2021reservoir,manuylovich2024}, which does not require the use of auxiliary delay lines.

In this work we present a novel method for optical RC without delay line, which exploits four-wave mixing assisted chaotic optical frequency combs (OFCs) in high $Q$-factor Kerr MR. Our approach (i) proposes for the first time to exploit modulation instability (MI) gain induced OFC generation to perform optical computing; (ii) it leverages intrinsic high-$Q$ MR memory capabilities bypassing the need of adding extra delay lines, which enhances computational speed; (iii) it enables nonlinear interaction between optical modes of the systems achievable for few tens of mW optical power injection to operate higher dimensionality transformation of input symbols necessary for computation, hence offering low power consumption. Optical frequency combs are ultraprecise optical rulers consisting of equally spaced coherent frequency laser lines that can usually be generated by several platforms, including mode-locked lasers \cite{Diddams}, electro-optic modulators \cite{Parriaux:20} and optical microresonators \cite{Pasquazi2018}. OFCs possess a variety of applications ranging from optical clocks to microscopy, from sensing to space exploration and telecommunications to mention just a few \cite{Fortier}. While for the vast majority of applications it is required that high stability and a fixed phase relation exist between the comb modes, in this work we leverage the potential technological application of less explored but easier to generate OFCs exhibiting random amplitude and phase fluctuations in their spectral lines.

\begin{figure*}
\includegraphics[scale=0.1]{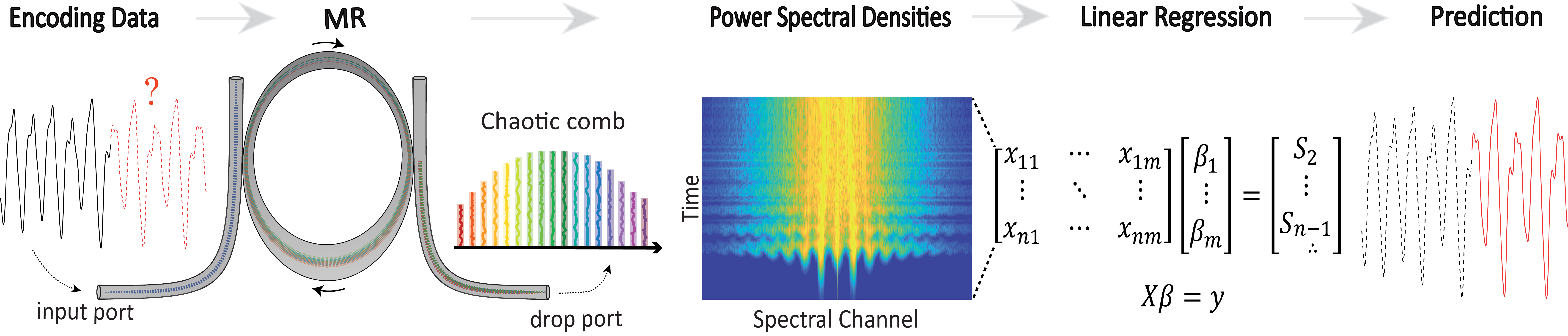}
\caption{\label{fig:overview_rc_model}Overview of the proposed RC model. The driving field encodes time-series information into the light wave which drives the MR. The input light then enters the MR. The MR high-dimensional output state consists of power spectral densities corresponding to a chaotic frequency comb. Linear regression is then applied to combine generated spectral features, then the next symbols in the chaotic time series are predicted.}
\end{figure*}
The proposed RC scheme operates as follows: a symbols time series generated from a chaotic dynamical system is encoded, after suitable rescaling, into input pump power values at which an optical high-Q factor Kerr MR is driven. The symbol input (driving pump power) to the resonator is updated at a rate of 1 GSa/s. The nonlinear dynamics of light inside the resonator, due to the MI process, induces a dramatic spectral broadening leading to the generation of a chaotic frequency comb corresponding to a large number of oscillating equally spaced cavity resonances with amplitudes and phases which randomly change in time (see Supplemental Material for more details). 
The readout layer consists of the power spectral densities of the MR individual resonances as they can be detected by an optical spectrum analyzer. 
After training the reservoir, we applied linear regression to combine the output features and predict the next symbol in the time series. In high-$Q$ resonators, the photon lifetime, given by $\tau_p=Q/\omega_0$ being $\omega_0$ the carrier frequency, can significantly exceed the inverse of the input symbol rate. This extended photon lifetime ensures that the spectral features generated by one input symbol persist in the cavity long enough to overlap with subsequent symbols enabling interaction of different symbols inside the cavity. This is further enhanced by the bistable response of Kerr cavities which is associated to an hysteresis cycle of the intracavity power as a function of the injection.
Figure \ref{fig:overview_rc_model} provides an overview of the proposed RC model, illustrating each step: from data encoding till time series prediction.
%To test our RC framework's capabilities and robustness, we validated our approach for prediction of time series generated from several benchmark chaotic dynamical systems, described by the Mackey-Glass, Rössler, and Lorenz equations.

%power spectral density of individual cavity resonances constitutes the n. 
%The rich nonlinear four-wave mixing dynamics o

%We use the driving field to encode time-series information \textcolor{pink}{(?)} into the light. The light then enters a MR made of silica with a refractive index of 1.44 and a cavity length of 0.001 mm. The output state consists of spectral channels, high-dimensional states represented by power spectral densities. A single input power yields power spectral densities at various frequencies, creating a high-dimensional nonlinear function. Key nonlinear features such as bistability and modulation instability (MI) generate new frequencies from an input continuous wave pump, thus enabling the nonlinear interactions necessary for complex computations. The modifiable output layer allows for easy training, enhancing adaptability and efficiency. We apply linear regression to combine these features and predict the next symbol in the time series. 
We modeled the dynamics of light inside the MR using the established mean field Lugiato-Lefever equation (LLE) in the notation of \cite{coen2013modeling}, which describes
the evolution over time \( t \) of the electric field slowly varying envelope \( E(t, \tau) \):
\begin{multline}\label{eq:LLE}
    \frac{\partial E}{\partial t} =  -\frac{(\alpha + i\delta_0)}{t_R} E - i\frac{L}{t_R} \frac{\beta_2}{2} \frac{\partial ^2 E}{\partial \tau^2}
    + i\frac{\gamma L}{t_R}|E|^2 E + \frac{\sqrt{\theta}}{t_R} E_{in}.
\end{multline}
Here, \( t_R \) represents the round-trip time of the cavity, \( \delta_0 \) is the detuning of the driving field from the nearest linear resonance, $L$ is the cavity length, $\beta_2$ the group velocity dispersion coefficient, $\gamma$ the nonlinearity coefficient, $E_{in}$ is the continuous wave pump injection amplitude related to the injected power by $P_{in}=|E_{in}|^2$.
\( \alpha \) and $\theta$ characterize cavity losses and coupling strength as a function of the resonator $Q$-factor, which is defined as $Q=\left(Q_{\text{int}}^{-1}+Q_{\text{ext1}}^{-1}+Q_{\text{ext2}}^{-1}\right)^{-1}$ and depends on the contributions from internal losses $Q_{int}$, losses associated to the input, $Q_{\text{ext1}}$, and drop, $Q_{\text{ext2}}$, waveguides respectively. To numerically solve Eq. \ref{eq:LLE} we employed a Fourth-Order Runge–Kutta Interaction Picture (RK4IP) method % \cite{hult2007fourth} 
with local tolerance control (see Supplemental Material for more details about parameters definitions and numerical methods).

\begin{figure}
\includegraphics[scale=0.2]{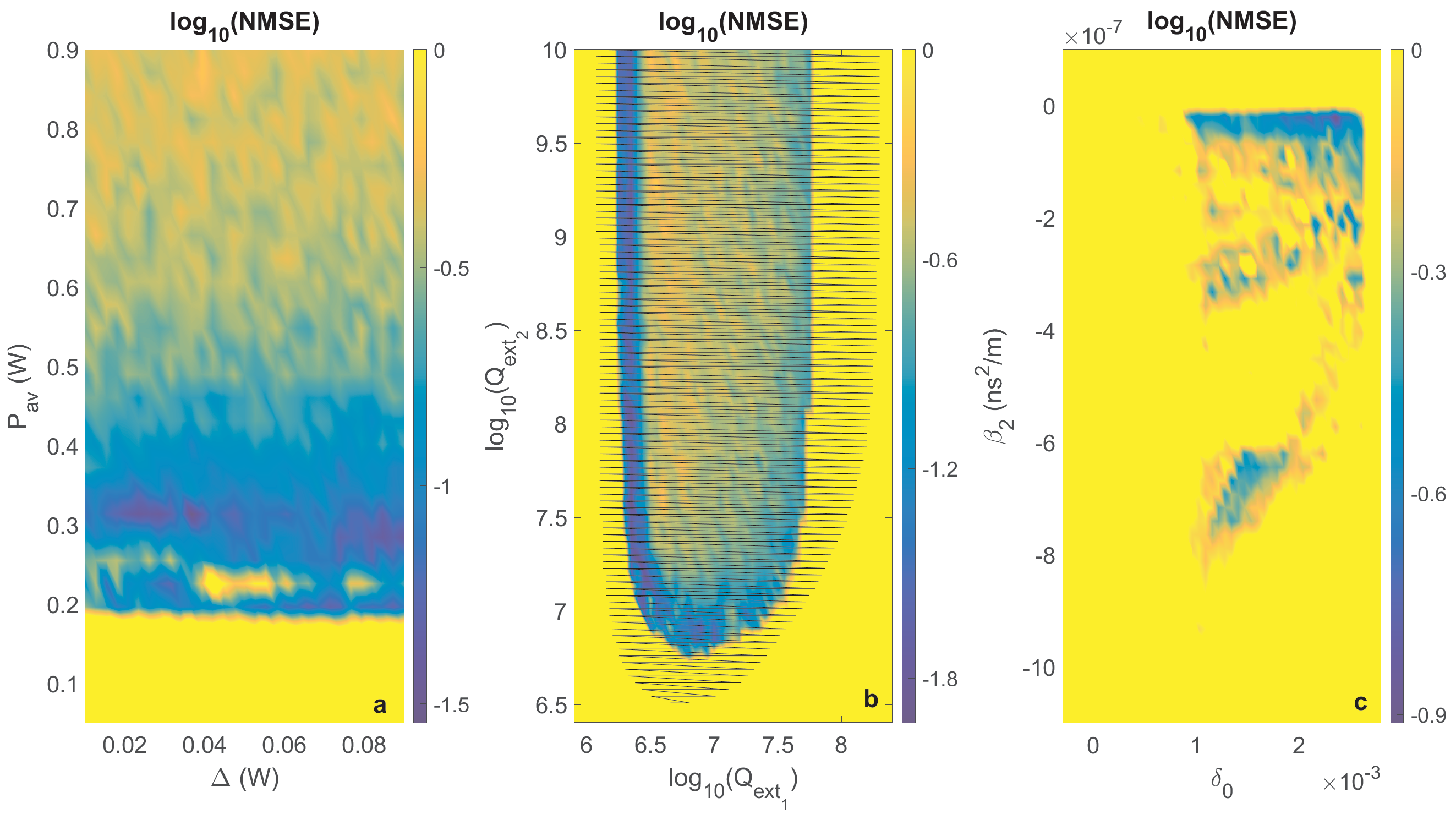}
\caption{\label{fig:MG_NMSE_datarate_1} 
(Mackey-Glass) Single-step prediction performance. (a) log\(_{10}\)(NMSE) as a function of average pump power (\(P_{av}\)) and pump power variation (\( \Delta \)). (b) log\(_{10}\)(NMSE) as a function of external quality factors \(Q_{\text{ext1}}\) and \(Q_{\text{ext2}}\). The shaded area shows the region where MI gain is larger than zero. (c)  log\(_{10}\)(NMSE) as a function of \( \beta_2 \) versus \( \delta_0 \). The colour scale indicates the NMSE, with lower values representing better prediction performance.
The default parameters are \(Q\)-factor \(= 1.0 \times 10^7\), \(\gamma = 1.6 \times 10^{-2}\) W$^{-1}$m$^{-1}$, \(\beta_2 = -4.0 \times 10^{-9}\) ns\(^2\)m$^{-1}$, \(\delta_0 = 1.6 \times 10^{-3}\), $\Delta=0.05$ W, $P_{av}=0.325$ W %\(P_{\text{min}} = 0.3\) W, \(P_{\text{max}} = 0.35\) W
, and the data rate is 1.0 GSa/s.}
\end{figure}

We tested the prediction capabilities of our MR based optical reservoir computer on 3 different tasks, and namely in the prediction of chaotic time series generated by different dynamical system: Mackey-Glass, Rössler and Lorenz equations.
We run simulations of these dynamical systems (see Supplemental Material for more details) generating for each one a time series featuring 1700 symbols.
We then mapped the time series into pump power values $P_{in}\in [P_{av}\pm\Delta/2]$ where $P_{av}$ is the average power and $\Delta$ defines the pump power variation, e.g. the excursion between maximum and minimum pump power injected. Such mapping could be experimentally implemented by temporally controlling the power of the continuous wave pump laser by means of a modulator. %such that the maximum of the series corresponded to $P_{max}=0.35$ W and the minimum to $P_{min}=0.3$ W.
For each value of the resonator input pump power, we built the RC feature matrix having as rows the power spectral densities collected from the MR drop port at regular time intervals -- in the ns scale (see individual picture captions) -- corresponding to the inverse of input symbol rate. The power spectral densities (as shown in the 'Power Spectral Densities' section of Fig. 1), representing the high-dimensional states generated by the MR, are hence transformed into features that form the input to the linear regression model. Each row of the feature matrix corresponds to a specific input symbol, and the matrix is then used to predict the next symbol in the time series.
To enhance stability and emphasize relevant features, the power spectral density sequence is first filtered around the pump frequency using a super-Gaussian filter of order 4, and a width of 1 THz. For each series we used $80\%$ of the data points for training and $20\%$ for testing, resulting in a feature matrix of $1360\times512$ with a rank typically of the order of $512$.
We then applied linear regression to estimate the optimal weights array $(\beta_1,...,\beta_m)$ (see Supplemental Material). 
When constructing the feature matrix, we also took into account the initial phase of intra-cavity energy build-up. This transient "warm-up" period was removed from the feature matrix to ensure that only quasi-steady-state characteristics of the system were included.
The performance of the RC system on each task was evaluated using a standard indicator and namely the Normalized Mean Squared Error (NMSE), which quantifies the prediction error (see Supplemental Material for the NMSE defintion). 
%\section{Results}
%\subsection{Single-step Prediction}
%The prediction methodology employed in this paper begins with the initial internal state of the MR, using the first symbol to generate features. Linear regression is then applied to combine these features and predict the next symbol. 
Figure \ref{fig:MG_NMSE_datarate_1} illustrates the single-step prediction performance as a function of the $Q$-factor, detuning $\delta_0$, and dispersion $\beta_2$. 
The colour scale represents the prediction error, with lower NMSE values indicating better prediction performance. 
The MI process requires a minimum amount of average power to start which translates into a clear threshold, $P_{av}\approx 0.2$ W,  needed to achieve low NMSE value as we can appreciate from Fig.\ref{fig:MG_NMSE_datarate_1} (a). Figure \ref{fig:MG_NMSE_datarate_1} (b) shows the computer performance as a function of external $Q$-factors ($Q_{\text{ext1}}$ and $Q_{\text{ext2}}$). We observe that the reservoir computer is capable of prediction ($\text{log}_{10}(\text{NMSE})<0$) when MI gain is different from zero. Hence MI is a necessary condition for the operation of our device. From Fig. \ref{fig:MG_NMSE_datarate_1} (b) we can also appreciate the existence of a region of optimal performance (dark blue stripe for $\text{log}_{10}(Q_{\text{ext1}})\approx 6.5$). We furthermore verified that this area moves to the left (lower values of $Q_{\text{ext1}}$ are needed to achieve optimal performance) when decreasing the input symbol rate. For input data rate slower than $\approx$ 0.1 GSa/s and for input data rate in the order of 10 GSa/s the device is not successful in performing prediction (see Supplemental Material for more details).
We can intuitively explain these limitations as follows: when input data rate is too slow the currently injected symbol dominates the dynamics and the memory of previous symbols is lost; on the other hand when input data rate is very high the new symbols do not have enough time to build up  and to leave their "imprint" inside the cavity. Hence an optimal input data rate exists which reaches a trade-off between these two effects.
In Fig. \ref{fig:MG_NMSE_datarate_1} (c) $\text{log}_{10}(\text{NMSE})$is shown as a function of group velocity dispersion $\beta_2$ and detuning $\delta_0$. We can qualitatively explain optimal performance of the system (blue areas corresponding to low NMSE in Fig.\ref{fig:MG_NMSE_datarate_1} (c)) as for low absolute value of anomalous dispersion the parametric gain spectrum is broader, which enables the generation of a larger number of spectral features. Narrow range of detuning values for which the performance is optimal can be interpreted as follows. For $\delta_0 \approx 0$ the cavity is monostable. By increasing the value of $\delta_0$ around $10^{-3}$, much larger average intracavity power can be reached due to the appearance of bistability in the cavity response: with relatively low injected pump power in the order of few hundreds mW the intracavity power can be in excess of 100 W, which can lead to broadband parametric gain and random OFC formation.
By increasing even further the value of the detuning -- $\delta_0 > 2.5\times 10^{-3}$ -- the upper branch of the cavity response is bistable for values of the injection much larger than those used in our simulations -- $P_{in}\approx 0.3$ W -- hence only the lower branch of the bistability curve can be accessed (see  Supplemental Material).
It is also important to stress that existence of MI is a necessary condition for the operation of our optical reservoir computer but it does not explain alone the parameter regions of optimal performance (dark blue regions in Fig. \ref{fig:MG_NMSE_datarate_1} (b)). We have performed the same characterization for time series generated from the Rössler and Lorenz system, and the results were similar to those obtained for the Mackey-Glass time series (see Supplemental Material).

%\subsection{Multi-step Prediction}
\begin{figure}
\includegraphics[scale=0.2]{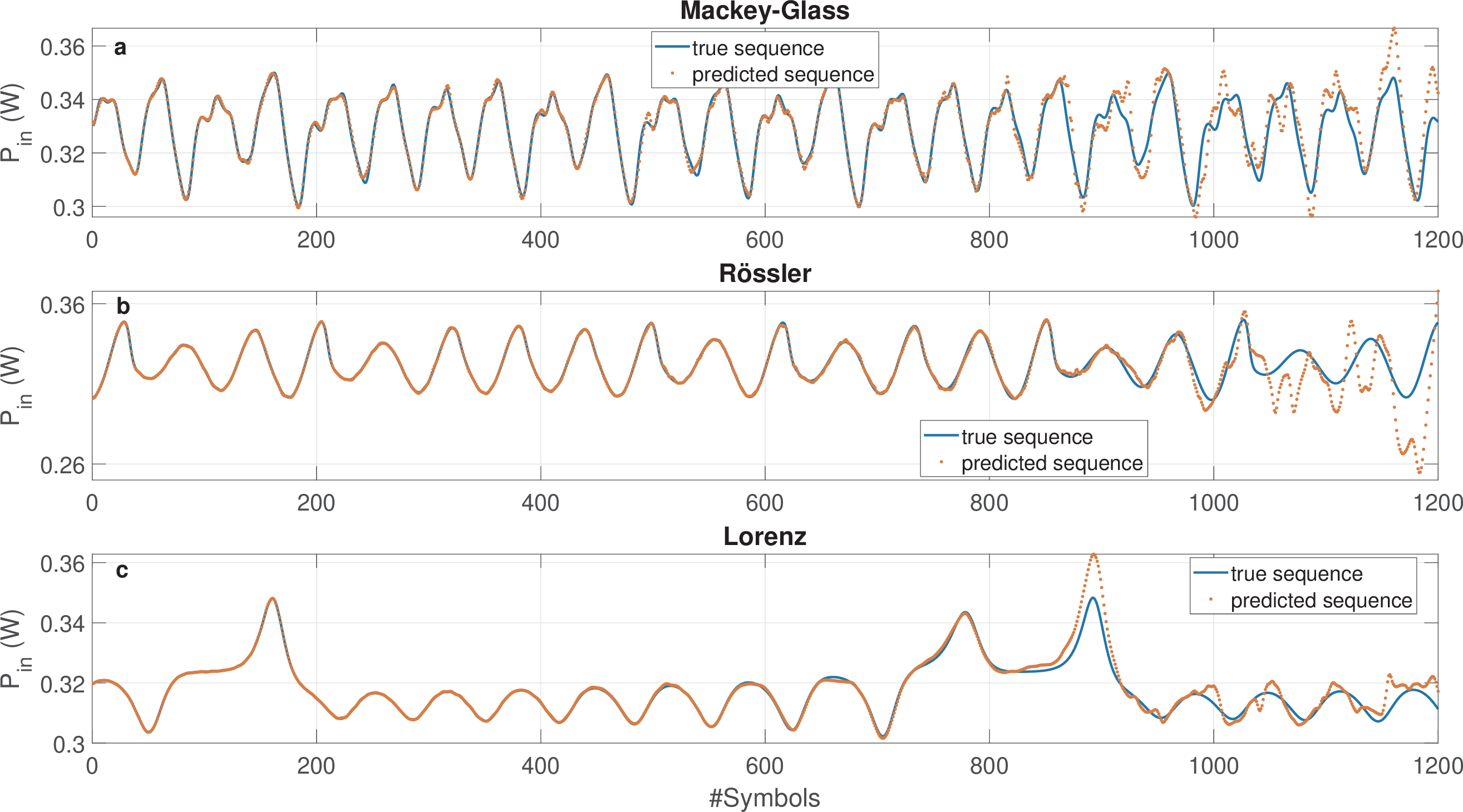}
\caption{\label{fig:multi_prediction} Examples of multi-step auto-regressive predictions using features generated from a single symbol for three benchmark dynamical systems: (a) Mackey-Glass time series, (b) Rössler attractor, and (c) Lorenz system. Each subplot shows the true sequence (blue line) alongside the predicted sequence (red dots) over 1200 symbols. Parameters are the same as in Fig. \ref{fig:MG_NMSE_datarate_1}.}
\end{figure}
We have also tested our RC in more advanced multi-step prediction tasks. We began with the same procedure as in single-step prediction, where the initial internal state of the MR is used to generate spectral features from the first symbol. These features are then used in a pseudoinverse linear regression model to predict the next symbol (input power). The predicted power is fed back into the MR to generate new features, continuing the cycle in an auto-regressive manner. This process allows the system to predict subsequent symbols, achieving predictions up to 1200 symbols ahead of a single input symbol.

Figure \ref{fig:multi_prediction} illustrates examples of multistep autoregressive prediction using features generated from a single symbol for the Mackey-Glass time series \ref{fig:multi_prediction} (a), Rössler  \ref{fig:multi_prediction} (b), and Lorenz system \ref{fig:multi_prediction} (c). The figure displays the true sequences (blue line) and the predicted sequences (red dots), showing that the predicted power values closely follow the true patterns, thereby demonstrating the accuracy and effectiveness of the approach. The regression method used is pseudoinverse, with 1700 symbols for training and testing and 700 symbols for warm-up.

\begin{figure}[!ht]
\includegraphics[scale=0.2]{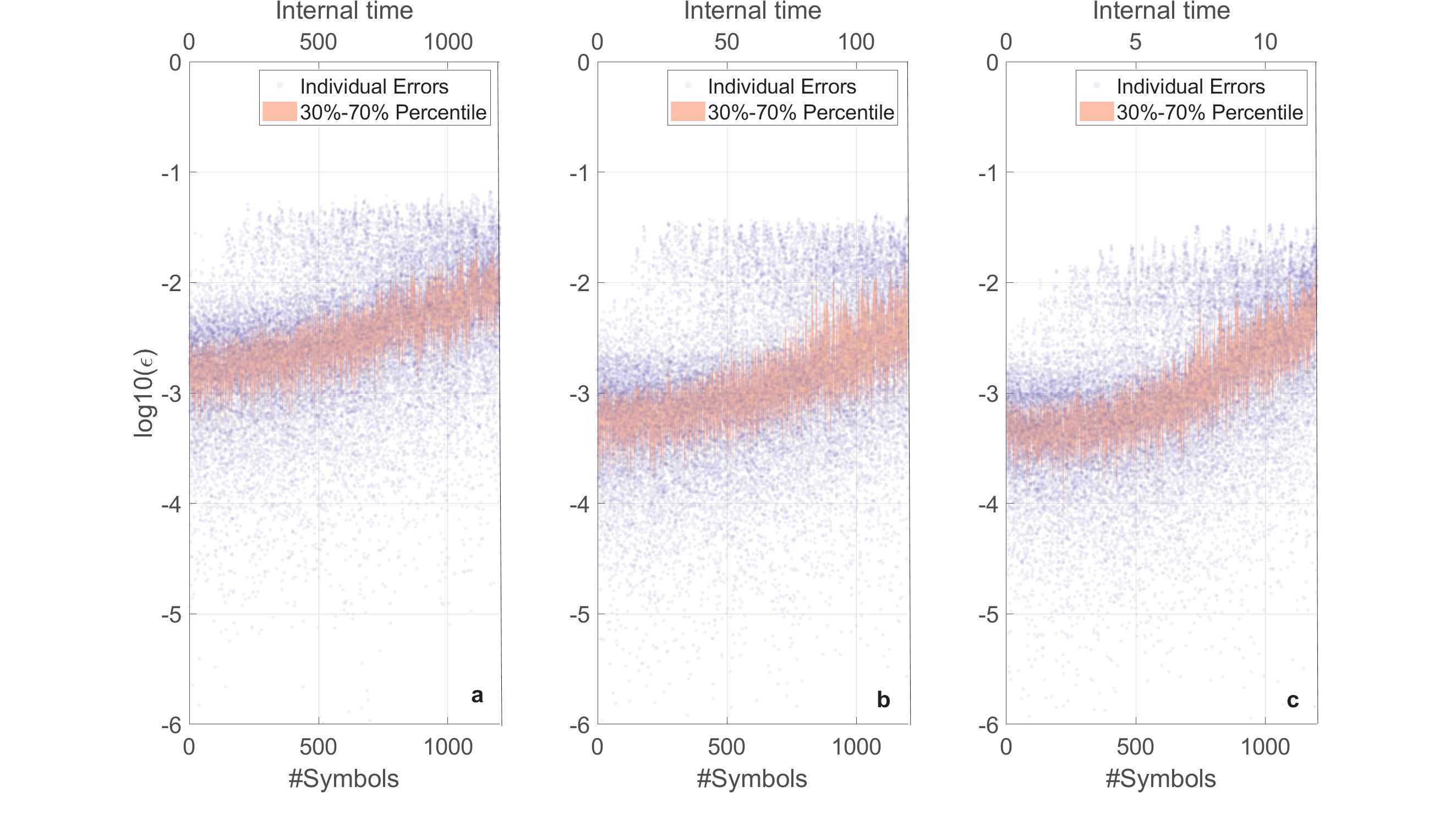}
\caption{\label{fig:error_analysis} Logarithmic error analysis of multi-step auto-regressive predictions for three benchmark dynamical systems: (a) Mackey-Glass time series, (b) Rössler attractor, and (c) Lorenz system. The plot shows the logarithmic scale of the absolute prediction errors, \(\log_{10}(\epsilon)\), for each of the 1200 symbols in the 20 predicted sequences. Individual errors are represented as purple dots, while the shaded pink region indicates the 30th to 70th percentile range of the errors, demonstrating the spread around the median error. Parameters used are the same as in Fig. \ref{fig:multi_prediction}.}
\end{figure}

In order to quantify the multi-step prediction capability of our RC, we calculated the RC error $\epsilon=|y_{\text{true}} - y_{\text{pred}}|$, where $y_{\text{true,pred}}$ stand for the true and predicted value respectively, for the multi-step auto-regressive predictions across 20 time-series, each obtained by seeding the dynamical systems (Mackey-Glass, Rössler, and Lorenz) with different input conditions. Results are reported in Fig. \ref{fig:error_analysis}. The plot shows the logarithm of the absolute prediction errors for each of the 1200 symbols in each of the 20 predicted sequences, with the individual errors for each sequence depicted as purple dots. The shaded pink region represents the 30th to 70th percentile range of the prediction errors, providing a measure of the spread of the errors around the median.
The $x$-axis of the plot has two time references. The lower axis, labeled as "\(\#\)Symbols", represents the "depth" of the autoregressive prediction, with each of the predicted symbols consecutively being encoded in the amplitude of light and transmitted to the MR. The upper axis, "Internal time", represents the corresponding time scale of evolution of the dynamical system that we use to test our RC. 
% underlying system's dynamics, scaled by the integration time step (\(dt\)). The relationship between these two axes is governed by a data rate of 1 GSa/s, meaning that each symbol represents 1 nanosecond of real experimental time, while the internal time accounts for the progression in terms of the simulated dynamics, specific to each dynamical system's \(dt\) parameter.
% The tight clustering of errors around the central percentile range indicates consistent predictive performance, with the majority of predictions falling within a narrow error margin. However, the occasional spread in the error dots suggests instances where the model struggles more with predictions, which might correspond to more chaotic or less predictable regions of the dynamical systems.

%The plot underscores the robustness of the model, as most prediction errors remain small (closer to 0 on the log scale), especially within the middle percentiles, demonstrating the model's ability to maintain accuracy over extended prediction horizons.

%\section{Discussion}
Our chaotic comb-based RC provides a solution for fast optical RC with excellent computational performance and a compact setup that exploits state-of-the-art integrated photonic devices.
One of its advantages, compared to many alternative MR-based architectures \cite{Donati:22,castro2023multi,GironCastro:24}, is the delay-free approach, which enables a compact architecture while also taking advantage of the high intrinsic resonator factor $Q$ for memory functionality. 
The potential computational speed of our device is in the GHz range exceeding a few hundred MHz reported in other MR-based RC computing approaches \cite{borghi2021reservoir}. Moreover, in our system, photon lifetime, which dictates memory, is directly influenced by the total $Q$-factor. For the range of $Q$-factors with low NMSE, the calculated photon lifetime ranges between 1.78 ns and 5.62 ns. Since the lowest photon lifetime exceeds the 1 ns sampling rate, this ensures that at least two symbols interact within the cavity, enabling effective memory retention and improving performance in tasks like time-series prediction.
Silicon MR RC schemes based on two-photon absorption and free carrier dispersion effects can operate at input power in the order of few mW due to the low threshold of the nonlinear dynamics exploited, however they have not demonstrated multiple symbols prediction capability to the best of our knowledge\cite{borghi2021reservoir,staffoli2023nonlinear, phang2020chaotic, bazzanella2022microring,Donati:22,GironCastro:24,castro2023multi}. 
While we have reported RC operation in the range of 0.2-0.3 W input pump power, it is worth stressing that in our simulations we have taken a conservative approach choosing a relatively small value of Kerr nonlinearity coefficient $\gamma$ compatible with fused silica MRs. Much larger values -- more than two orders of magnitude -- are used in the literature for MR made of materials such as silicon nitride or magnesium fluoride\cite{coen2013modeling} or others\cite{KipRev}. For larger values of $\gamma$ the MI process triggering spectral broadening can occur at lower input power.
%In our approach and for the parameters considered,  the necessary MI process required for spectral broadening a larger input power in the order of a few hundred mW, although we do not exclude that other effective computational performances can be achieved with input power of the order of mW for different Kerr MR architectures,for instance considering materials with a larger nonlinearity coefficient. 
In our work, we have also presented a novel, application of chaotic OFCs having non locked phases of neighbouring cavity modes. Chaotic comb states have so far mostly been avoided for applications due to their intrinsic instability with limited exceptions in the field of LIDARs \cite{ChaosCombs1,ChaosCombs2}.  
To conclude, we have proposed a novel method for delay-free optical reservoir computing based on chaotic frequency combs spontaneously generated by nonlinear four-wave mixing dynamics in a driven optical Kerr microresonator, demonstrating multi-step prediction of about 1000 consecutive symbols in chaotic time series with task-independent operation. Our results open new research directions for low-power consumption and low-footprint optical reservoir computing on a chip with ultrafast optical components.
\begin{acknowledgments}
AMP acknowledges support from the Royal Academy of Engineering through the Research Fellowship scheme.
EM acknowledges the support of the EU H2020 ITN project POSTDIGITAL (No. 860360) and the Engineering and Physical Sciences Research Council (project EP/W002868/1).
\end{acknowledgments}

\bibliography{apssamp}% Produces the bibliography via BibTeX.

\newpage
\onecolumngrid
\begin{center}
\Large    Supplemental Material
\end{center}
\section{Numerical methods}
The LLE has been solved numerically using an adaptive step 4th-order Runge-Kutta interaction picture (RK4IP) method \cite{4397001}.  Originally developed for the Generalized Nonlinear Schrödinger Equation (GNLSE), we modified the RK4IP method to simulate nonlinear dynamics in Kerr microresonators as described by the LLE. The simulation used 512 grid points, spaced by the resonator free spectral range (FSR) of 200 GHz; other simulation parameters are provided in the figure captions. 
Our notation for the LLE is the one given by \cite{coen2013modeling}, see \cite{Pasquazi2018} for reference to parameters conversion for coupled mode theory MR modelling.

The RK4IP method separates different terms of the LLE. For the LLE equation
\begin{equation}
\frac{\partial E}{\partial t} = -\frac{(\alpha + i\delta_0)}{t_R} E + i \frac{L}{t_R} \frac{\beta_2}{2} \frac{\partial^2 E}{\partial \tau^2} + i \frac{\gamma L}{t_R} |E|^2 E + \frac{\sqrt{\theta}}{t_R} E_{\text{in}}, \label{eq:LLE}
\end{equation}
we define the operator (\( L_{\text{op}} \)) which includes the dispersion detuning and loss terms:
  \[
  L_{\text{op}} = -\frac{(\alpha + i\delta_0)}{t_R} - i \frac{L}{t_R} \frac{\beta_2}{2} \frac{\partial^2}{\partial \tau^2},
  \]

and a second operator (\( N_{\text{op}}(E) \)) which incorporates the Kerr nonlinearity and the pump term:
  \[
  N_{\text{op}}(E) = i \frac{\gamma L}{t_R} |E|^2 E + \frac{\sqrt{\theta}}{t_R} E_{\text{in}}.
  \]

The algorithm for the RK4IP method proceeds in three main steps:

\textbf{Interaction Picture Transformation}: We define the field in the interaction picture, \(\hat{E}_{IP}\), by separating out the effect of the linear operator \(L_{\text{op}}\). The Fourier-transformed field \(\hat{E}_{IP}\) in the interaction picture is given by:
\[
\hat{E}_{IP}^{(n)} = \mathcal{F}\left[E^{(n)}\cdot \exp(-L_{\text{op}} \cdot t)\right],
\]
where \(\mathcal{F}\) represents the Fourier transform. 

%The original field \(E\) in the time domain can then be reconstructed from \(\hat{E}_{IP}\) by applying the inverse transformation:
%\[
%E^{(n)} = \mathcal{F}^{-1}\left[\hat{E}^{(n)}_{IP} \cdot \exp(L_{\text{op}} \cdot t)\right],
%\]
%where \(\mathcal{F}^{-1}\) is the inverse Fourier transform.

%We represent the field in the interaction picture (\(E_{IP}\)) by factorizing the exponential of the linear operator. The field \(E\)  in the time domain is expressed as:
%    \[
%    E^{(n)} = \mathcal{F}^{-1}\left[\hat{E}^{(n)}_{IP} \cdot \exp(L_{\text{op}} \cdot t)\right],
%    \]
%where \(\mathcal{F}^{-1}\) represents the inverse Fourier transform and \(\hat{E}_{IP}\) represents the Fourier transform of \(E_{IP}\).
 
\textbf{Nonlinear RK4 Step}:
The evolution of \(\hat{E}_{IP}\) is computed using a 4th-order Runge-Kutta method, applied only to the \(N_{\text{op}}(E)\) operator. The field \(\hat{E}_{IP}\) is evolved over a step \(\Delta t\) by calculating four intermediate slopes, \(k_1\), \(k_2\), \(k_3\), and \(k_4\), each evaluated at specific points in the interval \([t, t + \Delta t]\):

\begin{align*}
   k_1 &= N_{\text{op}}\left( t, \hat{E}_{IP}^{(n)} \right), \\
   k_2 &= N_{\text{op}}\left( t + \frac{\Delta t}{2}, \hat{E}_{IP}^{(n)} + \frac{\Delta t}{2} k_1 \right), \\
   k_3 &= N_{\text{op}}\left( t + \frac{\Delta t}{2}, \hat{E}_{IP}^{(n)} + \frac{\Delta t}{2} k_2 \right), \\
   k_4 &= N_{\text{op}}\left( t + \Delta t, \hat{E}_{IP}^{(n)} + \Delta t \, k_3 \right)
\end{align*}

The field \(\hat{E}_{IP}\) is then updated as:
\[
\hat{E}_{IP}^{(n+1)} = \hat{E}_{IP}^{(n)} + \frac{\Delta t}{6} \left( k_1 + 2k_2 + 2k_3 + k_4 \right)
\]
This weighted average of the slopes estimates the field at \( t + \Delta t \), with fourth-order accuracy in \(\Delta t\).

%The evolution of \(\hat{E}_{IP}\) is computed using a 4th-order Runge-Kutta method, applied only to the \(N_{\text{op}}(E)\) operator. The evolution is performed as:
 %  \[
  % \hat{E}^{(n+1)}_{\text{IP}} = \hat{E}^{(n)}_{\text{IP}} + \frac{\Delta t}{6} (k_1 + 2k_2 + 2k_3 + k_4),
   %\]
   %where the intermediate slopes \(k_1, k_2, k_3,\) and \(k_4\) are evaluated by applying \(N_{\text{op}}(E)\) at different stages in \(t\).

\textbf{Inverse Interaction Picture Transformation}: The field is then transformed back to the original frame by applying the inverse of the initial interaction picture transformation:
   \[
   E^{(n+1)} = \mathcal{F}^{-1}\left[\hat{E}^{(n+1)}_{\text{IP}} \cdot \exp(L_{\text{op}} \cdot t)\right],
   \]
   reconstructing the physical field and including both linear and nonlinear effects accumulated over the propagation step. Here \(\mathcal{F}^{-1}\) denotes the inverse Fourier transform.
    
\section{Q-factor}

The total quality factor \(Q\) of an MRR is determined by both intrinsic and extrinsic losses. The intrinsic Q-factor \(Q_{\text{int}}\) accounts for material absorption within the resonator, while the extrinsic $Q$-factors \(Q_{\text{ext1}}\) and \(Q_{\text{ext2}}\) correspond to losses at the coupling regions of the input and output ports, respectively. The overall $Q$-factor is expressed as:

\[
Q = \left(\frac{1}{Q_{\text{int}}} + \frac{1}{Q_{\text{ext1}}} + \frac{1}{Q_{\text{ext2}}}\right)^{-1}.
\]

The loss coefficient \(\alpha\) is linked to \(Q\) through the photon lifetime \(t_{\text{ph}}\), with \(\alpha = \frac{t_R}{2 t_{\text{ph}}}\), where \( t_R \) represents the round-trip time of the cavity, and \(t_{\text{ph}} = \frac{Q}{2 \pi \nu_0}\) which \(\nu_0\) represents the pump frequency. The coupling coefficients \(\theta_1\) and \(\theta_2\) are related to the extrinsic Q-factors \(Q_{\text{ext1}}\) and \(Q_{\text{ext2}}\), respectively \cite{Pasquazi2018}. Specifically, \(\theta_1\) is given by \(\theta_1 = \frac{t_R}{t_{e1}}\), with \(t_{e1} = \frac{Q_{\text{ext1}}}{2\pi \nu_0}\), and \(\theta_2\) is given by \(\theta_2 = \frac{t_R}{t_{e2}}\), where \(t_{e2} = \frac{Q_{\text{ext2}}}{2\pi \nu_0}\).
So the output of the resonator from the drop waveguide, which is where the power spectral densities used for prediction tasks are taken, is given by $E_2=\theta_2E$.

The introduction of the drop waveguide, whose contribution to the overal MR $Q$-factor is given by \(Q_{\text{ext2}}\), enables to extract the intracavity spectrum avoiding strong superposition with the reflected injection occuring at the input waveguide. 
 Without this auxiliary waveguide, the output power from the input port would primarily reflect the driven field entering the waveguide. 

\section{Modulation instability gain}
We summarize the analytical expression for the Modulation Instability (MI) gain derived from the linear stability analysis of the stationary solution of the LLE (Eq.\ref{eq:LLE}) used to study the behaviour of light in an MR. The MI gain in a nonlinear optical system describes the growth of small perturbations on top of a steady-state continuous wave solution (\(E_s\)). In the presence of small perturbations \(a(t, \tau)\) the field is expressed as \(E(t, \tau) = E_s + a(t, \tau)\), with $|a|<<|E_s|$. The MI gain describes how these perturbations evolve, depending on factors like nonlinearity, dispersion, and cavity detuning, which determines whether the perturbations grow (leading to instability) or decay. The eigenvalues  \(\lambda_\pm\), which characterize the stability of the continuous wave solution, can be obtained from linear stability analysis and are given by

\[
\lambda_{\pm} = \frac{J + J^*}{2} \pm \sqrt{\left(\frac{J - J^*}{2}\right)^2 + \gamma^2 L^2 |E_s|^4},
\]

with \(J\) and \(J^*\) (representing complex of \(J\)) depending on system parameters as:

\[
J = -(\alpha + i\delta_0) + iL\frac{\beta_2}{2}\omega^2 + 2i\gamma L |E_s|^2,
\]
 The power gain $G = 2\text{Re}(\lambda_+)$.

The steady-state solution \(E_s\) is found by relating the intracavity power \(P_s = |E_s|^2\) to the input power \(P_{\text{in}} = |E_{\text{in}}|^2\) through the following expression:

\[
P_{\text{in}} = \frac{\gamma^2 L^2 P_s^3 - 2 \gamma \delta_0 L P_s^2 + ( \alpha^2 + \delta_0^2) P_s}{\theta_1}.
\]

%where \(P_s = |E_s|^2\) represents the explicit form of the stationary solution.

\section{Cavity Response}

\begin{figure}[!h]
\includegraphics[scale=0.3]{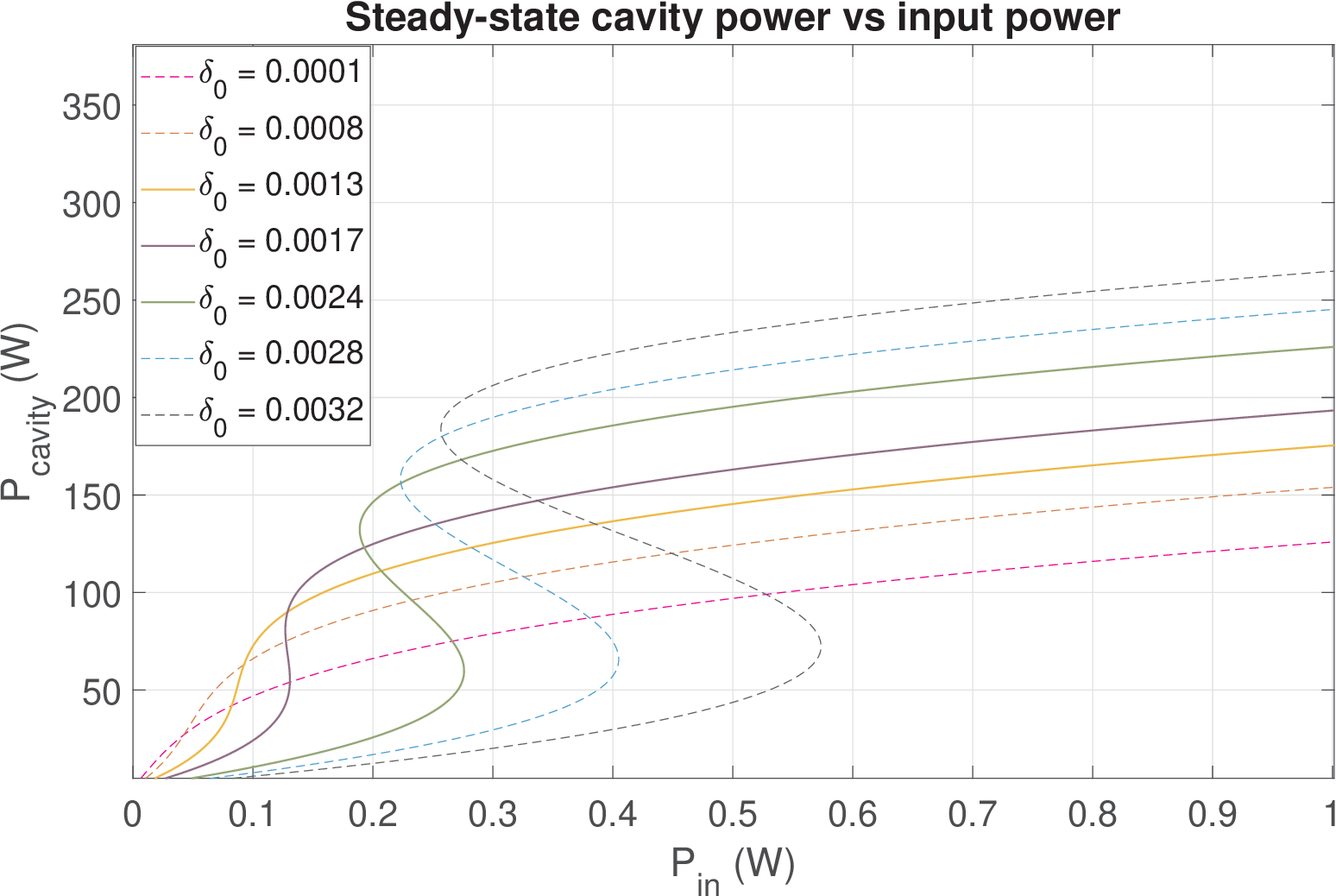}
\par Fig. S1. Intracavity power as a function of input power for different values of the cavity detuning. Parameters used are: \(Q\)-factor \(= 1.0 \times 10^7\), \(\gamma = 1.6 \times 10^{-2}\) W$^{-1}$km$^{-1}$.%\(\beta_2 = -4.0 \times 10^{-9}\) ns\(^2\)/m 
\label{fig:bistability}
\end{figure}
In Fig. S1 the cavity response -- intracavity power -- is plotted as function of the input pump power $P_{in}$ for several values of the cavity detuning $\delta_0$. Note that for the S-shaped cases the negative slope branch is always unstable. The continuous lines indicate solutions for which the upper branch of the bistability curve can be reached with moderate injection below 0.3 W. Reaching a large intracavity power with relatively low input is necessary to obtain substantial spectral broadening which enables effective operation of the reservoir computer. Hence an understanding of the bistable cavity response is instrumental in an effective design of the proposed platform.

\section{Modulation Instability Gain}
In this section we report plots of the MI gain as a function of $Q$-factors, detuning and dispersion. In Fig. S2 the gain of the maximally growing frequency mode is plotted versus external $Q$-factors ($Q_{ext_1}$,$Q_{ext_2}$) (note that in Fig.2b of the main paper, the positive gain is plotted with shaded area overimposed to NMSE results). In Fig. S3 we plot the frequency dependent gain as a function of cavity detuning and group velocity dispersion.

\begin{figure}[!h]
\includegraphics[scale=0.2]{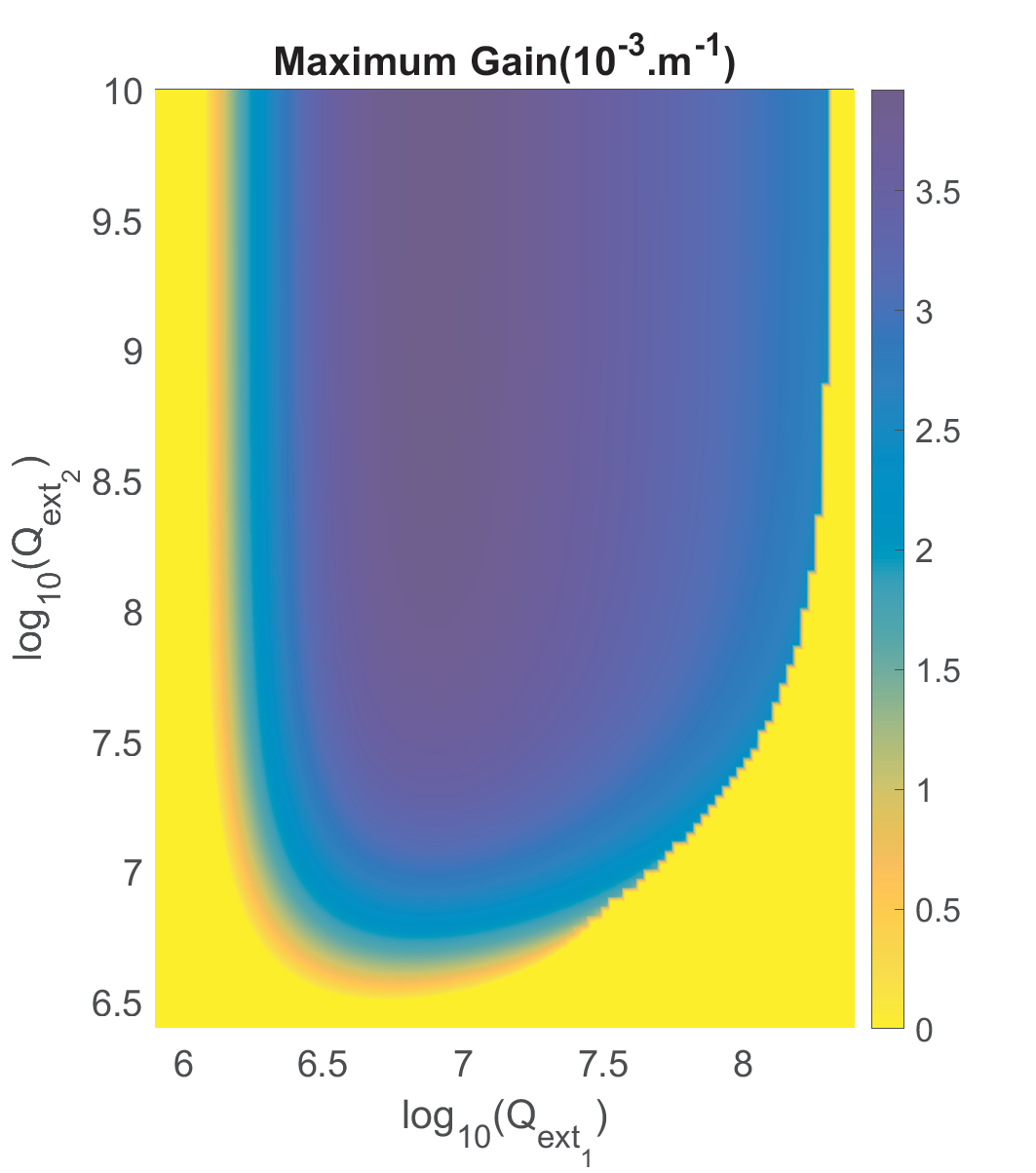}
\label{fig:maxGain} \par Fig. S2. Maximum MI gain is plotted as a function of external quality factors \(Q_{\text{ext1}}\) and \(Q_{\text{ext2}}\). Parameters: \(\gamma = 1.6 \times 10^{-2}\) W$^{-1}$m$^{-1}$, \(\beta_2 = -4.0 \times 10^{-9}\) ns\(^2\)m$^{-1}$, \(\delta_0 = 1.6 \times 10^{-3}\), $\Delta=0.05$ W.
\end{figure}
\begin{figure}[!h]
\includegraphics[scale=0.2]{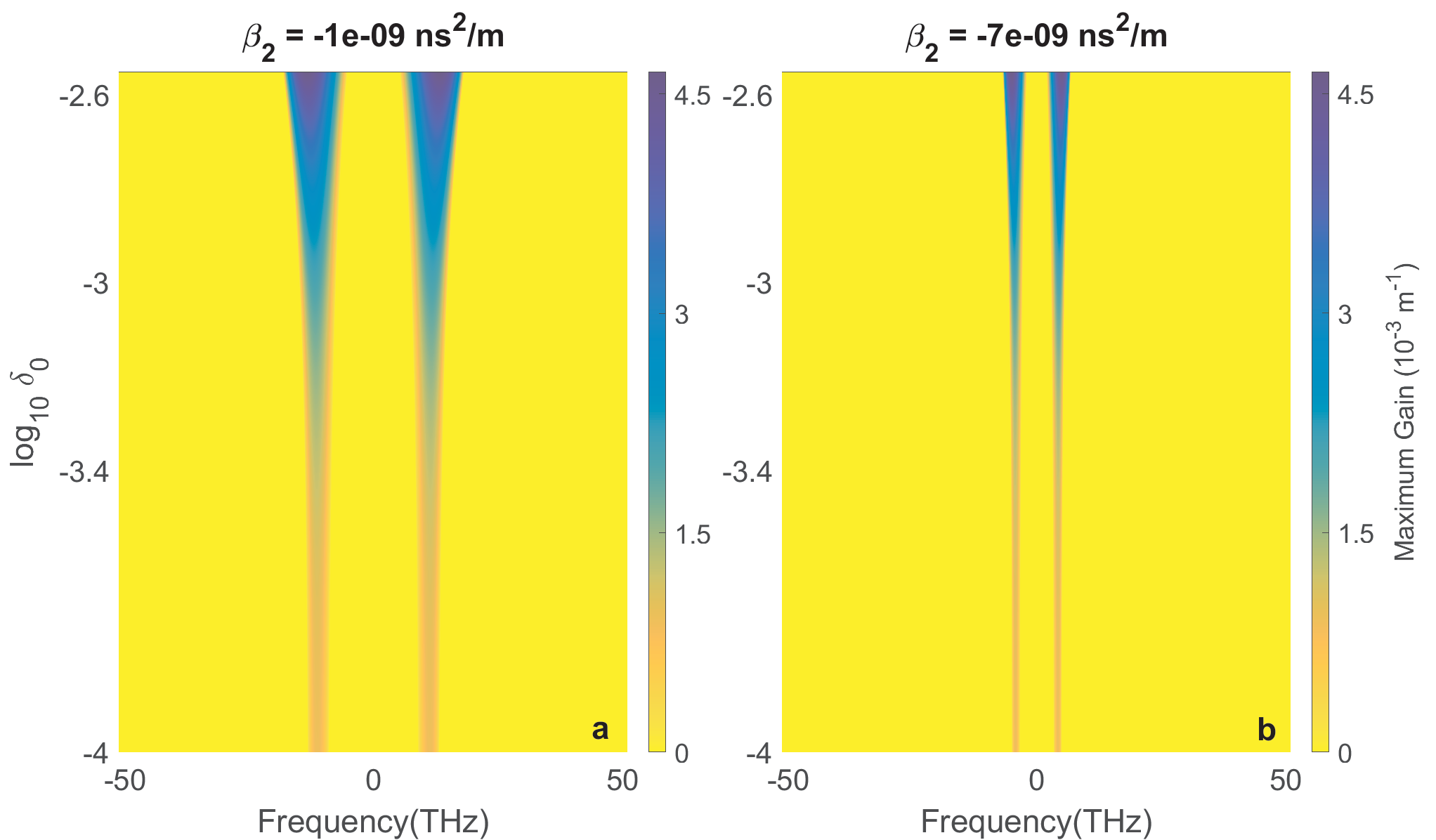}
\label{fig:beta2} \par Fig. S3. Maximum MI gain is plotted as a function of cavity detuning and group velocity dispersion. Remaining parameters are the same as in Fig. S2. %Parameters used are: \(Q\)-factor \(= 1.0 \times 10^7\), \(\gamma = 1.6 \times 10^{-2}\) 1/W/m, \(P_{\text{min}} = 0.3\) W, \(P_{\text{max}} = 0.35\) W.

\end{figure}

\newpage

\section{Random Frequency comb example}
In Fig. S4 we present an example of the spectral dynamics at fixed pump power for a constant input power in the range used for computation. Random fluctuations in the power spectrum are clearly visible in panel (a), while the temporal dynamics shown in panel (b) presents an unstable temporal pattern featuring pulse oscillations and collisions.
\begin{figure}[!h]
\includegraphics[scale=0.3]{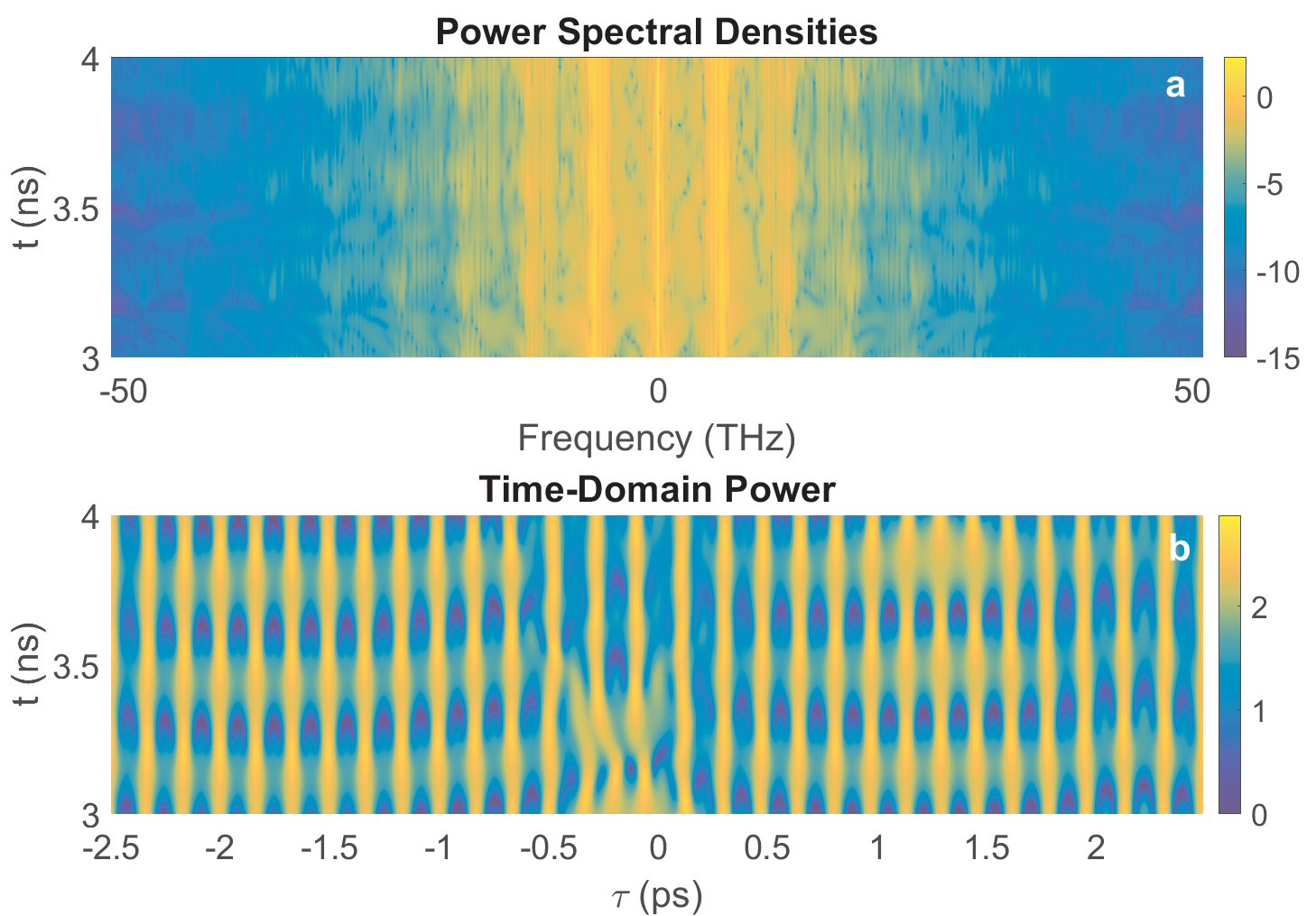}
\label{fig:PSDs_TPDs} \par Fig. S4. Evolution of power spectral densities for the constant input power \(P_{\text{in}} = 0.34\) W. The other parameters are the same as Fig. S2.
%Parameters used are: \(Q\)-factor \(= 1.0 \times 10^7\), \(\gamma = 1.6 \times 10^{-2}\) W$^{-1}$m$^{-1}$, \(\beta_2 = -4.0 \times 10^{-9}\) ns\(^2\)m$^{-1}$, \(\delta_0 = 1.6 \times 10^{-3}\)
\label{randomcomb}
\end{figure}

\section{Chaotic time series for prediction}
In order to obtain  chaotic time series to be used for prediction tasks 
we simulated three different dynamical systems: Mackey-Glass, Lorenz, and Rössler. 
\newline\newline
\textbf{Mackey-Glass system}

The Mackey-Glass system is a delay differential equation introduced to model physiological processes. The equation is:

\begin{equation}
\frac{dx(t)}{dt} =  \frac{\beta_{M} x(t-\tau_{M})}{1 + x(t-\tau_{M})^n_{M}} - \gamma_{M} x(t)
\end{equation}

where \( x(t) \) is the state variable at time \( t \), and \( \beta_{M} \), \( \gamma_{M} \), \( \tau_{M} \), and \( n_{M} \) are system parameters. In our simulations, we used the following parameter values: \( \beta_{M} = 0.2 \), \( \gamma_{M} = 0.1 \), \( \tau_{M} = 17 \), and \( n_{M} = 10 \).
\newline\newline
\textbf{Lorenz system}

The Lorenz system consists of the following three coupled first-order nonlinear differential equations:
\begin{align}
\frac{dx}{dt} &= \sigma_L(y - x) \\
\frac{dy}{dt} &= x(\rho_L - z) - y \\
\frac{dz}{dt} &= xy - \beta_L z
\end{align}

where \( x, y, z \) represent the state of the system, and \( \sigma_L, \rho_L, \beta_L \) are parameters. To simulate chaotic dynamics we used the following parameter values: \( \sigma_L = 10 \), \( \rho_L = 28 \), and \( \beta_L = \frac{8}{3} \).
\newline\newline
\textbf{Rössler system}

The Rössler system consists of three nonlinear ordinary differential equations:

\begin{align}
\frac{dx}{dt} &= -y - z \\
\frac{dy}{dt} &= x + a_R y \\
\frac{dz}{dt} &= b_R + z(x - c_R)
\end{align}

where \( x, y, z \) are the dynamical variables, \( t \) is time, and \( a_R, b_R, c_R \) are parameters that control the behavior of the system. In our simulations, we used the following parameter values \( a_R = 0.2 \), \( b_R = 0.2 \), and \( c_R = 5.7 \); which give rise to chaotic behavior.
\section{Linear regression}
\textbf{Training and estimation of weight matrix (\(\beta\))}

In this study, a linear regression model is employed to map input features to target outputs by estimating a weight matrix \(\beta\). The training process seeks to minimize the difference between the predicted outputs and the actual target values while incorporating regularization to mitigate overfitting.
Given a feature matrix \(\mathbf{X} \in \mathbb{R}^{n \times m}\), where \(n\) denotes the number of observations and \(m\) represents the number of features, and a target matrix \(\mathbf{Y} \in \mathbb{R}^{n \times k}\), where \(k\) is the number of target variables (with \(k=1\) in the case of single-symbol prediction), the weight matrix \(\beta \in \mathbb{R}^{m \times k}\) is estimated by minimizing the following regularized loss function:

\[
\mathcal{L}(\beta) = \frac{1}{2} \|\mathbf{Y} - \mathbf{X}\beta\|_F^2 + \frac{\lambda}{2} \|\beta\|_F^2,
\]

where \(\|\cdot\|_F\) denotes the Frobenius norm, and \(\lambda\) is a regularization parameter controlling the balance between model complexity and fit.

The optimal \(\beta\) is obtained using the closed-form solution:

\[
\beta = (\mathbf{X}^\top \mathbf{X} + \lambda \mathbf{I})^{-1} \mathbf{X}^\top \mathbf{Y},
\]

where \(\mathbf{I}\) is the identity matrix. This equation effectively regularizes the model by penalizing large coefficients, thus preventing overfitting.
\newline\newline

\textbf{Single-step prediction}

Once \(\beta\) has been estimated, it is used for single-step prediction. Given a new feature matrix \(\mathbf{X}_{\text{new}} \in \mathbb{R}^{p \times m}\), where \(p\) is the number of new observations, the predicted target values \(\mathbf{Y}_{\text{pred}} \in \mathbb{R}^{p \times k}\) are computed as:

\[
\mathbf{Y}_{\text{pred}} = \mathbf{X}_{\text{new}} \beta.
\]

This matrix equation performs a linear transformation of the new features using the learned weight matrix \(\beta\), yielding the corresponding predictions.
\newline\newline

\textbf{Multi-step prediction}

In multi-step sequential prediction, the model iteratively predicts a sequence of future values, where each prediction is based on the previously predicted values.

Starting with an initial feature matrix \(\mathbf{X}_0\), the first prediction is made as:

\[
\mathbf{Y}_1 = \mathbf{X}_0 \beta.
\]

The feature matrix is then updated by incorporating the new prediction \(\mathbf{Y}_1\) (e.g., by shifting and appending \(\mathbf{Y}_1\) to the feature set) to form \(\mathbf{X}_1\). The next prediction is made as:

\[
\mathbf{Y}_2 = \mathbf{X}_1 \beta.
\]

This process is repeated for \(d\) steps, generating a sequence of predictions \(\mathbf{Y}_1, \mathbf{Y}_2, \ldots, \mathbf{Y}_d\). Formally, for each time step \(t\) (\(t = 0, 1, \ldots, d-1\)):

\[
\mathbf{Y}_{t+1} = \mathbf{X}_t \beta,
\]

where \(\mathbf{X}_t\) is the feature matrix updated after each prediction to include the most recent predicted value. This iterative application of \(\beta\) enables the model to generate predictions over multiple future time steps.
\section{NMSE definition}
We quantified the capability of prediction of our system through the normalised mean squared error (NMSE) indicator. The NMSE is defined as 
\[ \text{NMSE} = \frac{\sum_{n} (y_{\text{true}, n} - y_{\text{pred}, n})^2}{\sum_{n} (y_{\text{true}, n} - \bar{y}_{\text{true}})^2},\]
where \( y_{\text{true}} \) is the symbol true value, \( y_{\text{pred}} \) is the symbol predicted value, and \( \bar{y}_{\text{true}} \) is the mean of the true values. The lower the NMSE value is the more accurate the prediction performance is.
%\section{Singular value decomposition}
%In Fig.\ref{SVD} we show an example of singular value decomposition of a typical power spectral density features matrix, showing that the number of singular values larger than XX and hence significantly contributing to prediction is YY. This represent a typical quantitative measure of the number of meaningful nonlinear features generated by the MR reservoir computer.
%\textcolor{red}{Negar, Egor, can you please expand on this adding numbers and improving my comments...?}
%\begin{figure}[!h]
%\includegraphics[scale=0.6]{SVD_PSDs.eps}
%\caption{\label{fig:PSDs_TPDs} (Mackey-Glass) Parameters: \(Q_{\text{int}} = Q_{\text{ext1}} = Q_{\text{ext2}} = 1.0 \times 10^7\), \(\gamma = 1.6 \times 10^{-2}\) 1/W/m, \(\beta_2 = -4.0 \times 10^{-9}\) ns\(^2\)/m, \(\delta_0 = 1.6 \times 10^{-3}\), \(P_{\text{min}} = 0.3\) W, \(P_{\text{max}} = 0.35\) W.}\label{SVD}
%\end{figure}
\section{Impact of Symbol Rate}
In the main paper we reported results obtained for an input data rate of 1 GSa/s. In this section we report results about lower and higher data rate.
We can see that with 0.1 GSa/s prediction performance are still excellent, while going to 10 GSa/s does not provide good performance. If input data rate is much smaller than 0.1 GSa/s the performance will drop as well.
\subsection{Data Rate  = 0.1 GSa/s}
\begin{figure}[!h]
\includegraphics[scale=0.22]{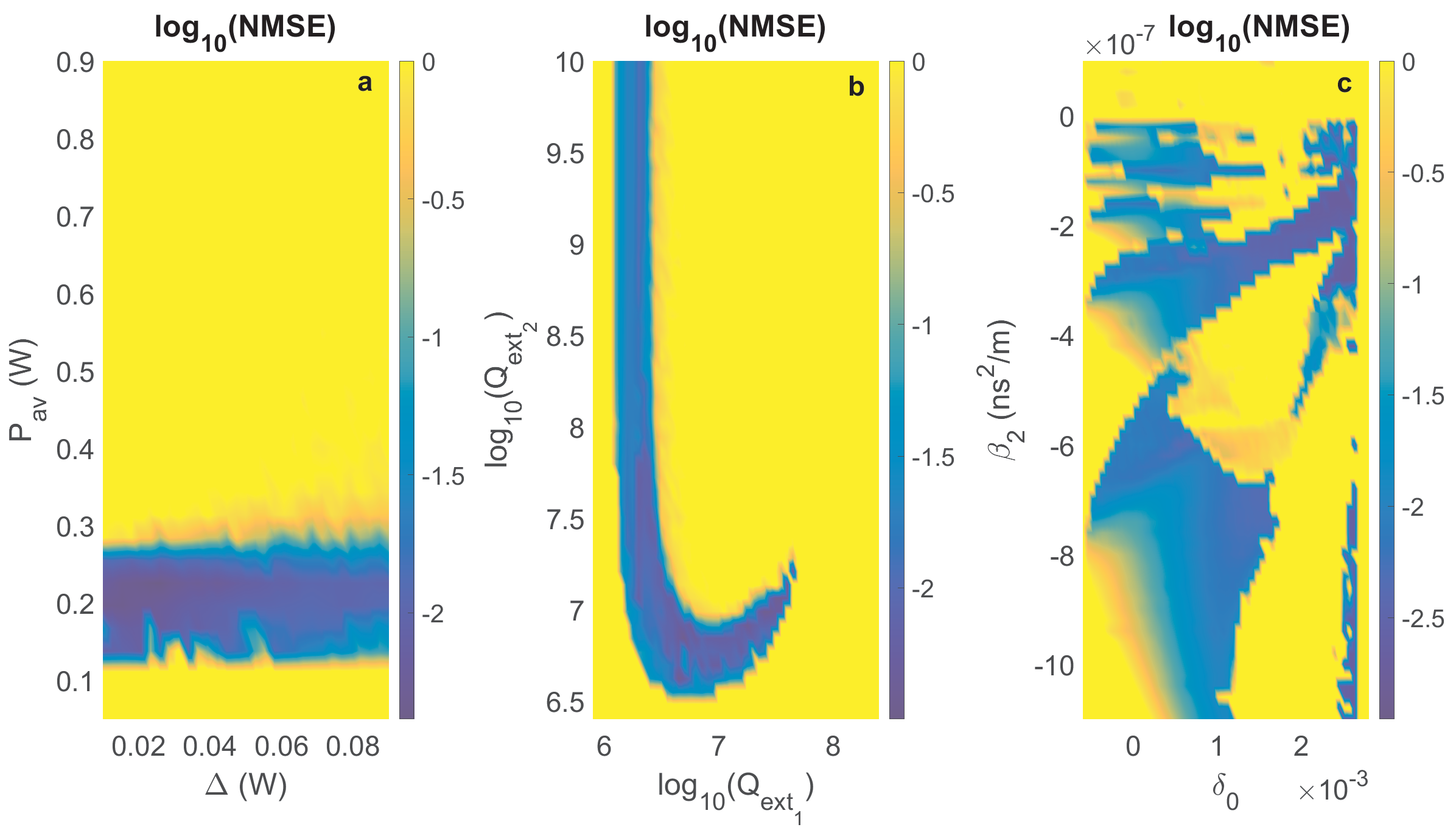}
\label{fig:MG_NMSE_datarate_01} \par Fig. S5. Single-step prediction performance for Mackey-Glass time series with an input data rate of 0.1 GSa/s. (a) log\(_{10}\)(NMSE) as a function of average pump power (\(P_{av}\)) and pump power variation (\( \Delta \)). (b) log\(_{10}\)(NMSE) as a function of external quality factors \(Q_{\text{ext1}}\) and \(Q_{\text{ext2}}\). (c)  log\(_{10}\)(NMSE) as a function of \( \beta_2 \) versus \( \delta_0 \). The default parameters are: \(Q\)-factor \(= 1.0 \times 10^7\), \(\gamma = 1.6 \times 10^{-2}\) W$^{-1}$m$^{-1}$, \(\beta_2 = -4.0 \times 10^{-9}\) ns\(^2\)m$^{-1}$, \(\delta_0 = 1.6 \times 10^{-3}\), \(P_{\text{min}} = 0.3\) W, \(P_{\text{max}} = 0.35\) W. The colour scale indicates the NMSE, with lower values representing better prediction performance.
\end{figure}
\clearpage
\subsection{Data Rate = 10 GSa/s}
\begin{figure}[!h]
\includegraphics[scale=0.22]{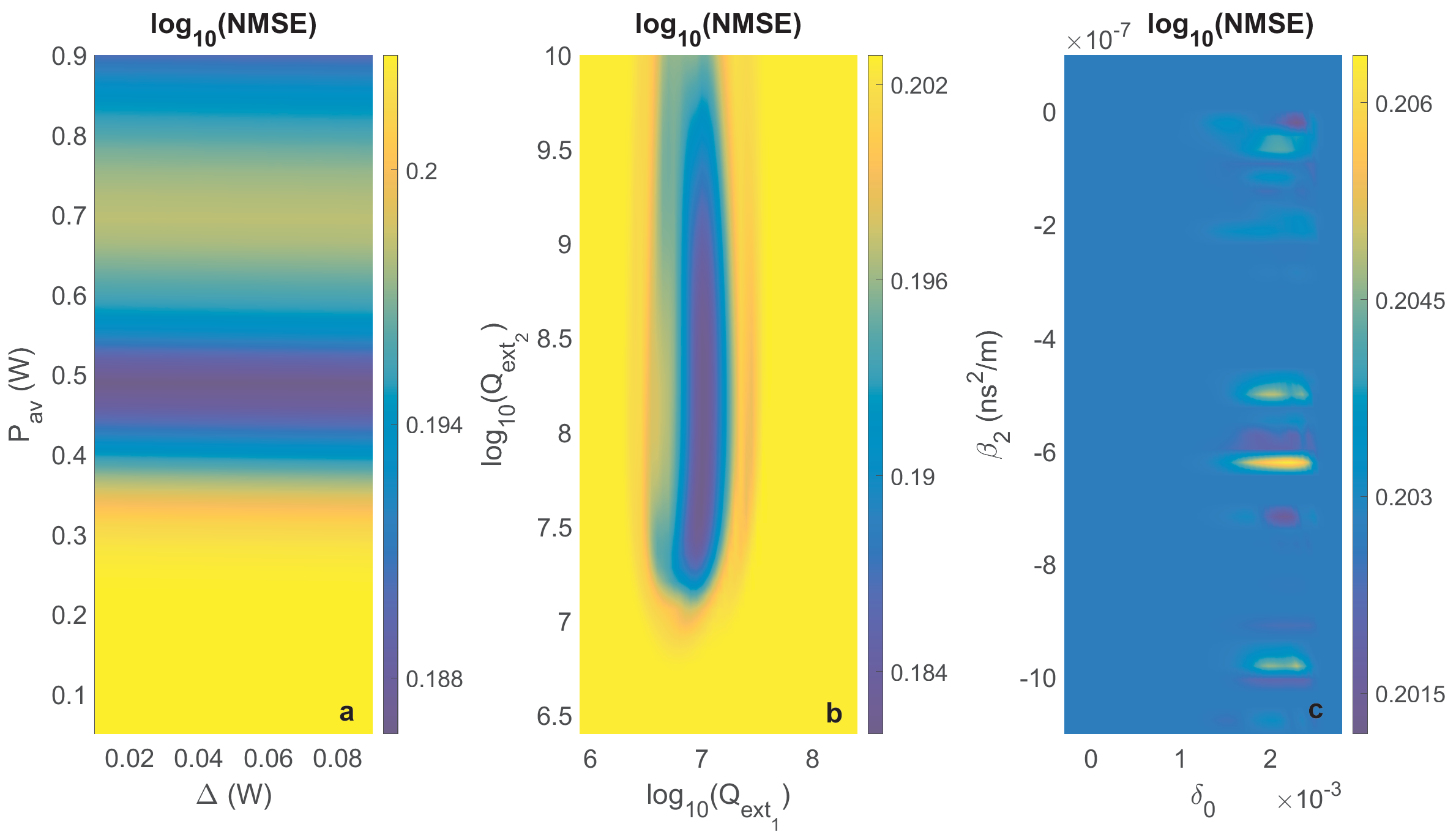}
\label{fig:MG_NMSE_datarate_10} \par Fig. S6. Single-step prediction performance for Mackey-Glass with input data rate of 10 GSa/s. (a) log\(_{10}\)(NMSE) as a function of average pump power (\(P_{av}\)) and pump power variation (\( \Delta \)). (b) log\(_{10}\)(NMSE) as a function of external quality factors \(Q_{\text{ext1}}\) and \(Q_{\text{ext2}}\). (c)  log\(_{10}\)(NMSE) as a function of \( \beta_2 \) versus \( \delta_0 \). Parameters are the same as Fig. S5.
\end{figure}

\section{Other tasks}
In Fig.2 of the main paper, we reported plots of the NMSE for Mackey-Glass time series prediction tasks as a function of several system parameters such as input power, $Q$-factors, cavity detuning and dispersion. In this section we report similar plots for Rössler and Lorenz system.

\subsection{Rössler System}
\begin{figure}[!h]
\includegraphics[scale=0.22]{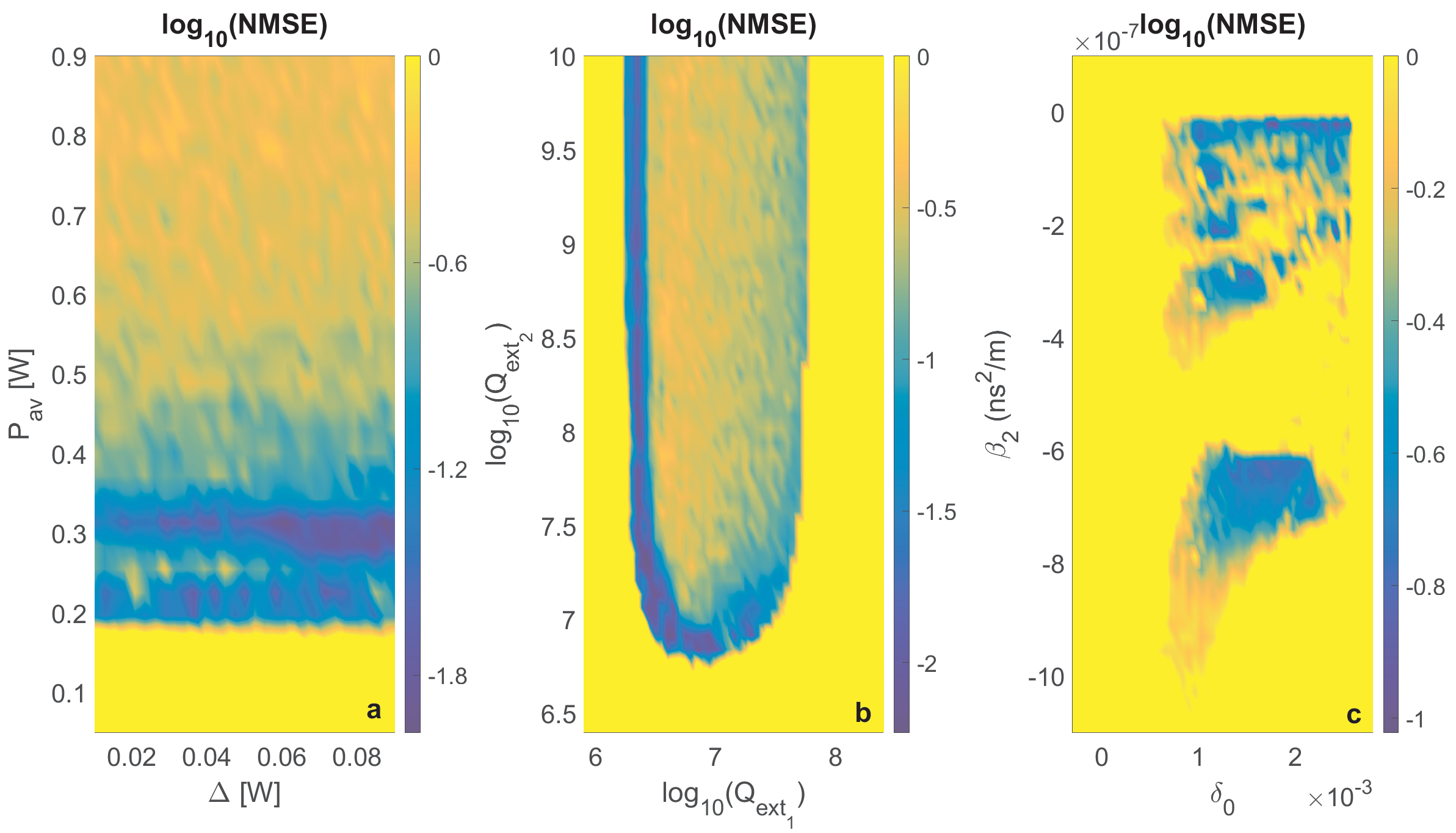}
\label{fig:Rössler_NMSE}
 \par Fig. S7. (Rössler) Single-step prediction performance. (a) log\(_{10}\)(NMSE) as a function of average pump power (\(P_{av}\)) and pump power variation (\( \Delta \)). (b) log\(_{10}\)(NMSE) as a function of external quality factors \(Q_{\text{ext1}}\) and \(Q_{\text{ext2}}\). (c)  log\(_{10}\)(NMSE) as a function of \( \beta_2 \) and \( \delta_0 \). The colour scale indicates the NMSE, with lower values representing better prediction performance. The data rate is 1.0 GSa/s. The other parameters are the same as Fig. S5.
\end{figure}\clearpage
\subsection{Lorenz System}
\begin{figure}[!h]
\includegraphics[scale=0.22]{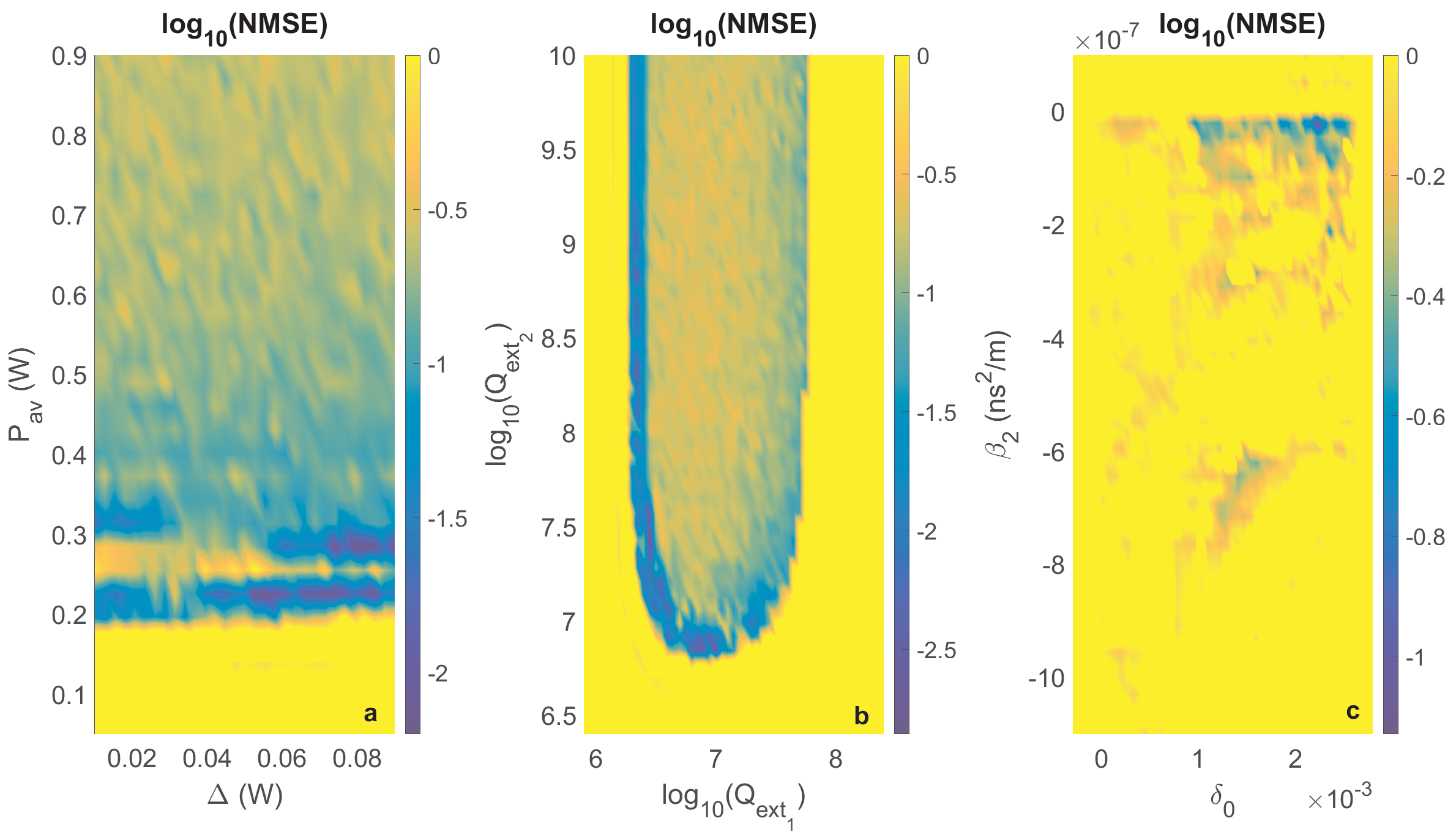}
\label{fig:Lorenz_NMSE} \par Fig. S8. (Lorenz) Single-step prediction performance. (a) log\(_{10}\)(NMSE) as a function of average pump power (\(P_{av}\)) and pump power variation (\( \Delta \)). (b) log\(_{10}\)(NMSE) as a function of external quality factors \(Q_{\text{ext1}}\) and \(Q_{\text{ext2}}\). (c)  log\(_{10}\)(NMSE) as a function of \( \beta_2 \) and \( \delta_0 \). The colour scale indicates the NMSE, with lower values representing better prediction performance. The data rate is 1.0 GSa/s. The other parameters are the same as in Fig. S5.
\end{figure}

\section{MASE}
In this section we report results about another indicator used in the community known as Mean Absolute Scaled Error (MASE) indicator.
The MASE is defined as 
\[\text{MASE} = \frac{\frac{1}{n} \sum_{n} |y_{\text{true}} - y_{\text{pred}}|}{\frac{1}{n-1} \sum_{n} |y_{\text{true}} - y_{\text{true,lagged}}|},
\]
where \( y_{\text{true}} \) is the true symbol value, \( y_{\text{pred}} \) is the predicted symbol value, and \( y_{\text{true,lagged}} \) represents the lagged true values used for scaling. The lower the MASE value, the more accurate the prediction performance is.

\subsection{Mackey-Glass}
\begin{figure}[!h]
\includegraphics[scale=0.22]{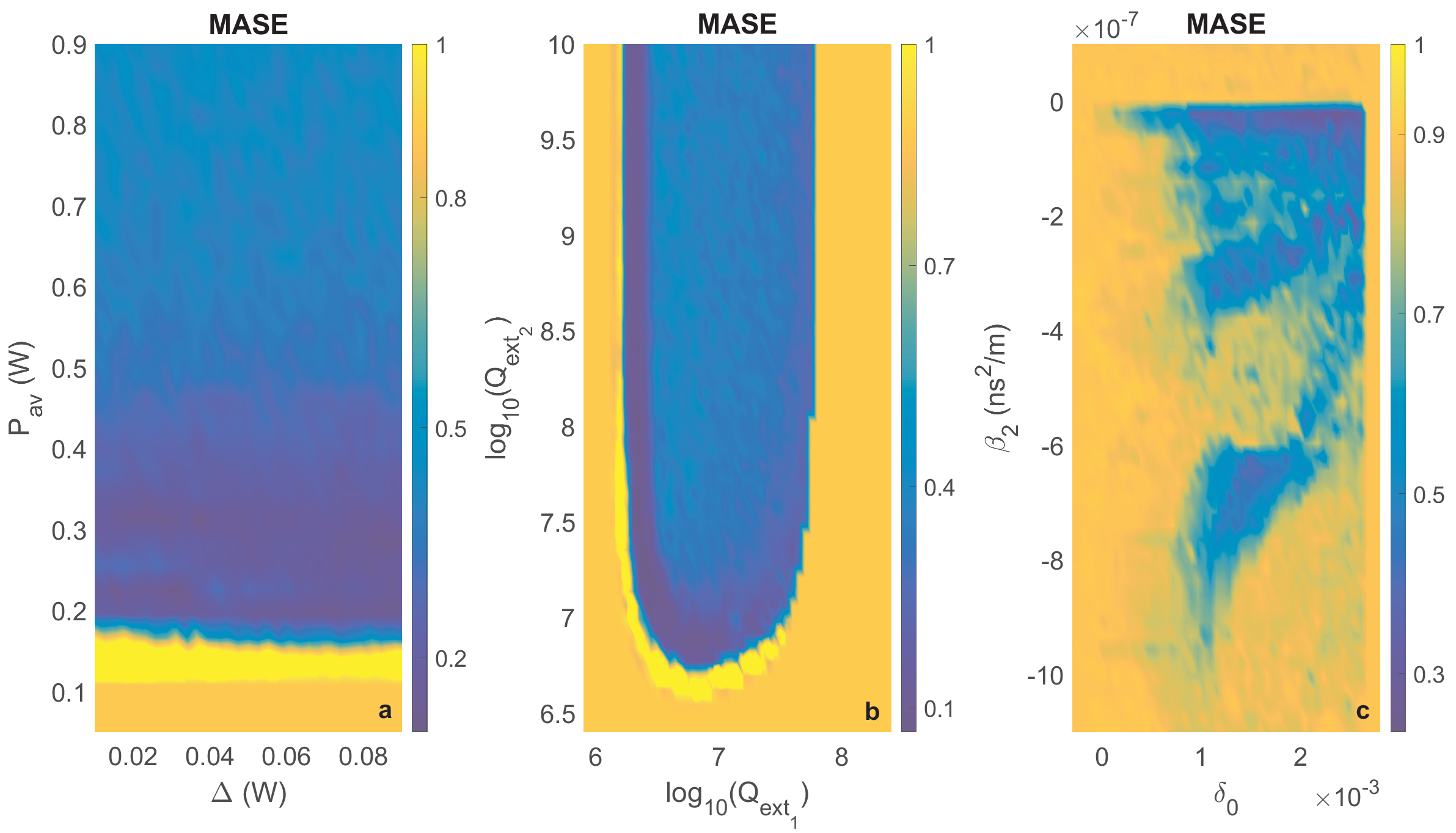}
\label{fig:MG_MASE} \par Fig. S9. (Mackey-Glass) Single-step prediction performance. (a) log\(_{10}\)(MASE) as a function of average pump power (\(P_{av}\)) and pump power variation (\( \Delta \)). (b) log\(_{10}\)(MASE) as a function of external quality factors \(Q_{\text{ext1}}\) and \(Q_{\text{ext2}}\). (c)  log\(_{10}\)(MASE) as a function of \( \beta_2 \) and \( \delta_0 \). The colour scale indicates the MASE, with lower values representing better prediction performance. The data rate in the 3 plots is 1.0 GSa/s. The other parameters are the same as in Fig. S5.
\end{figure}\clearpage
\subsection{Rössler}
\begin{figure}[!h]
\includegraphics[scale=0.22]{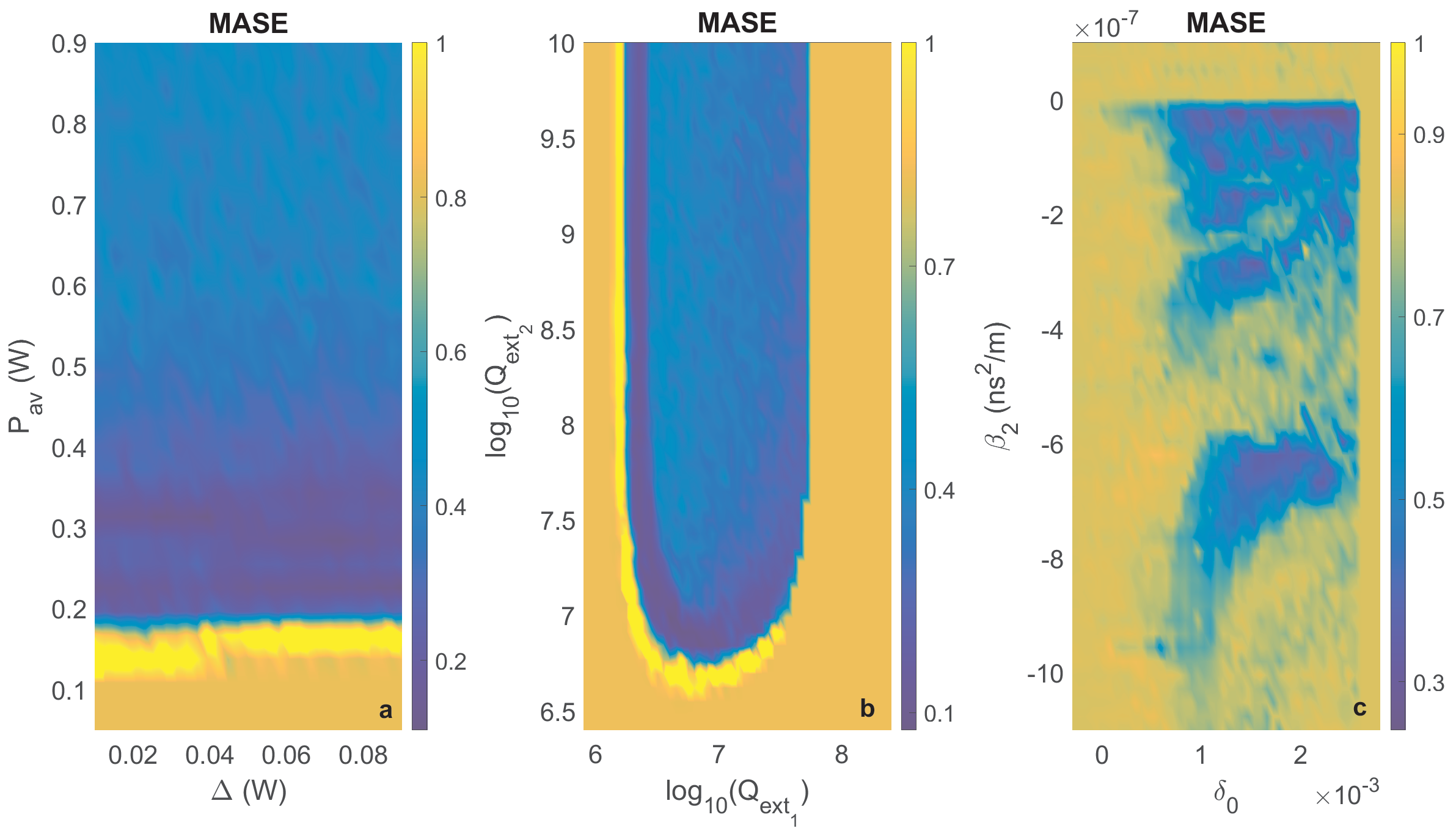}
\label{fig:Rössler_MASE} \par Fig. S10. (Rössler) Single-step prediction performance. (a) log\(_{10}\)(MASE) as a function of average pump power (\(P_{av}\)) and pump power variation (\( \Delta \)). (b) log\(_{10}\)(MASE) as a function of external quality factors \(Q_{\text{ext1}}\) and \(Q_{\text{ext2}}\). (c)  log\(_{10}\)(MASE) as a function of \( \beta_2 \) versus \( \delta_0 \). The colour scale indicates the MASE, with lower values representing better prediction performance. The data rate in the 3 plots is 1.0 GSa/s. The other parameters are the same as in Fig. S5.
\end{figure}

\subsection{Lorenz}
\begin{figure}[!h]
\includegraphics[scale=0.22]{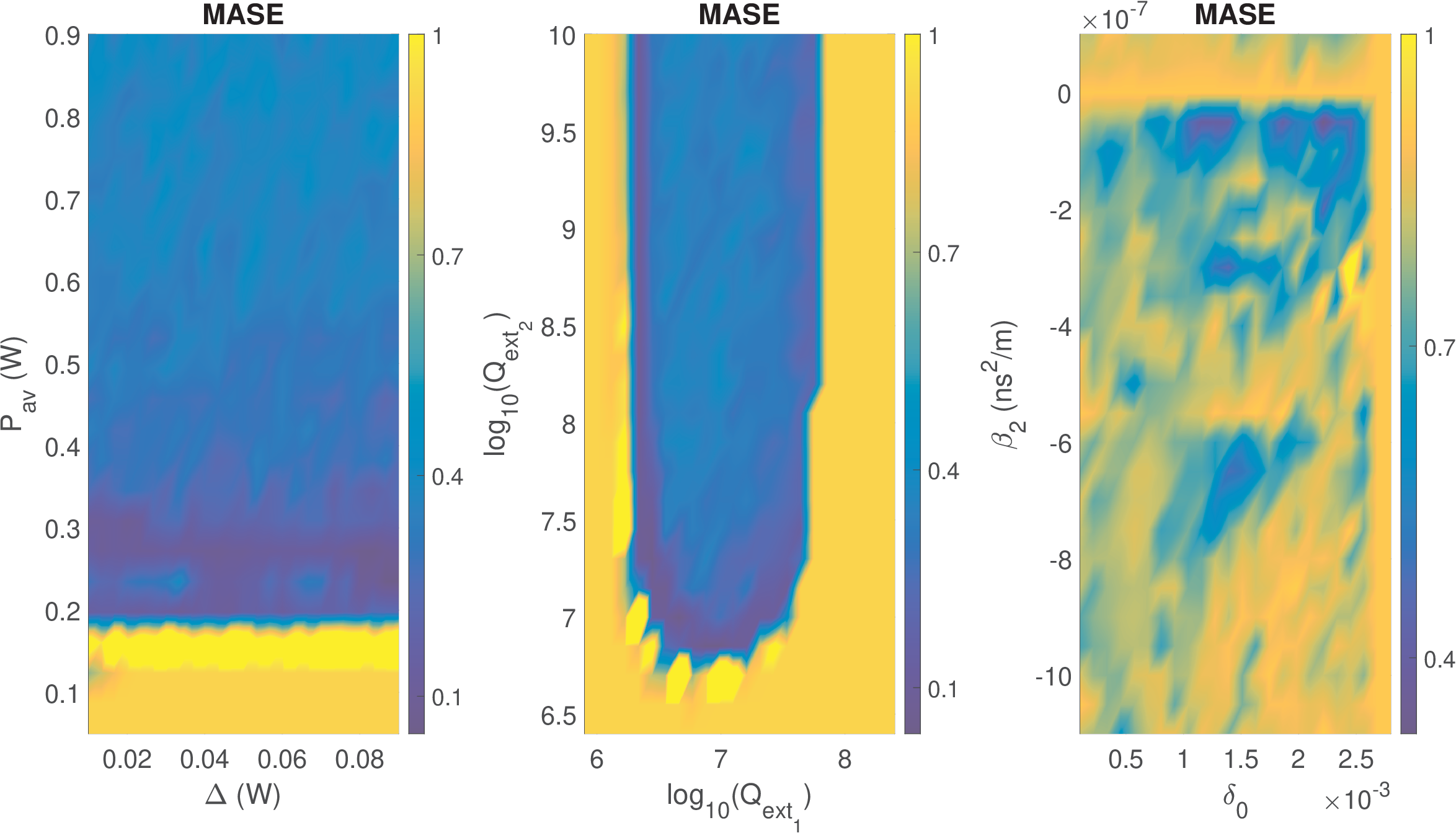}
\label{fig:Lorenz_MASE} \par Fig. S11. (Lorenz) Single-step prediction performance. (a) log\(_{10}\)(MASE) as a function of average pump power (\(P_{av}\)) and pump power variation (\( \Delta \)). (b) log\(_{10}\)(MASE) as a function of external quality factors \(Q_{\text{ext1}}\) and \(Q_{\text{ext2}}\). (c)  log\(_{10}\)(MASE) as a function of \( \beta_2 \) versus \( \delta_0 \). The colour scale indicates the MASE, with lower values representing better prediction performance. The data rate in the 3 plots is 1.0 GSa/s. The other parameters are the same as in Fig. S5.
\end{figure}

%\clearpage
%\nocite{*}
%\bibliography{apssamp}

\end{document}